# Charge coupling to anharmonic lattice excitations in a layered crystal at 800K.

F M Russell, Dept. of Physics, University of Pretoria, Pretoria 0002, Rep. South Africa.
7/05/2015    mica2mike@aol.com

Abstract. A study of charged particle and quasi-particle anharmonic lattice excitations in a meta-stable crystal of layered structure at high temperature has shown that a mobile lattice excitation called a quodon can bind securely to a hole or electron. Quodons are decoupled from phonons and can propagate along atomic chains of more than $10^8$ atoms at 800K. Their relevance to high Tc superconductors is discussed.

All high Tc superconductors reported so far have a layered crystal structure showing symmetry about the central [001]-plane[1]. Here the possible relevance of anharmonic phenomena is considered. The interaction of nonlinear lattice excitations with single electronic charges in layered crystals is difficult to study experimentally due to the inability to identify the location, type and energy of the excitations in a solid. The fortuitous inclusion of impurities in natural crystals of the mineral muscovite, which has a layered structure somewhat resembling that of high Tc materials, has enabled the creation and propagation of charges bound to nonlinear excitations to be probed in some detail. The main finding is that *anharmonic* lattice excitations can bond securely to single electronic charges of either sign at temperatures exceeding 800K and propagate at near sonic speed for distances exceeding $10^8$ unit cells in muscovite.

The unit cell of muscovite $KAl_2(AlSi_3O_{10})(F,OH)_2$ has a layered structure consisting of a flat octahedrally coordinated mono-atomic sheet of potassium sandwiched between two identical layers of linked $(Si,Al)O_4$ tetrahedra. Large crystals grow in liquid granite at temperatures above 800°K at depths exceeding 5km. Most crystals contain isomorphous atomic replacements including Ca for K and Fe or Mg for Al as well as random impurities. During initial cooling, while migration is still possible, crystals that depart significantly from stoichiometric composition become structurally meta-stable. In this state perturbations of the lattice can initiate irreversible phase changes resulting in local formation of different minerals, principally magnetite $Fe_2O_3$ and epidote $Ca_2(Al,Fe)_3(SiO_4)_3(OH)$. At a depth of 5km the main source of perturbations is local radioactivity with a small contribution of muons from cosmic rays. Potassium is radioactive, emitting mostly neutrinos, electrons, gamma rays and occasionally positrons. The lattice instability is charge sensitive, positive charge allowing magnetite and negative charge allowing epidote to form. This leads to the tracks of positrons and positive muons being delineated with black magnetite, allowing them to be observed and identified optically[2,3].

Potassium has five decay channels: about 90% of which give electrons, 10% gammas and 0.0001% positrons, all with neutrinos. Each decay causes the nucleus to recoil with maximum energies of 10 eV for positrons and 42eV for most electrons[4]. The angular distribution of positron tracks in the (001)-plane of potassium, where the track recording process operates, is determined by diffraction scattering. This causes the recoiling atom to move in the direction of atomic chains lying in principal crystallographic directions. The interaction of an energetic recoiling atom with adjacent atoms in the chains is strongly nonlinear and creates mobile lattice excitations, called quodons. Model studies of such interactions in 1D chains and 2D arrays indicated that the energetic mobile excitations are probably intrinsic localised modes called breathers[5]. At the high energies involved in quodon creation by K-decay numerical modelling indicates the excitations are optical-mode breathers. Confirmation of this awaits more powerful molecular dynamic programs. Supporting evidence for quodons was provided by a laboratory experiment in which one edge of a crystal of muscovite held at 300K was bombarded with alpha particles. Quodons created near the surface in atomic cascades then propagated along atomic chains for more than $10^7$ unit cells before ejecting atoms from the rear edge of the crystal by inelastic scattering[6].

The emission of an electron leaves a positive charge at the point of creation of the quodon. The initial energy of these quodons is variable as it depends on the distribution of energies and momenta between the electron, neutrino and the recoil nucleus. Measurements of the width of the tracks of quodons as they propagate show that the average amount of magnetite deposited per unit length of track is constant over the length of the track and is the same for other quodon tracks. This shows that the recording process is not determined primarily by the kinetic energy of a quodon. The variation of width of positron tracks as a function of distance from the stopping point is shown in Figure 1. It increases as the particle slows down reaching a broad maximum when it is moving near sonic speed. Also plotted is the mean width for quodon tracks, which matches closely that for a positive charge moving at near sonic speed. This result is consistent with a quodon consisting of a breather moving at slightly sub-sonic speed bound to a positive charge or hole.

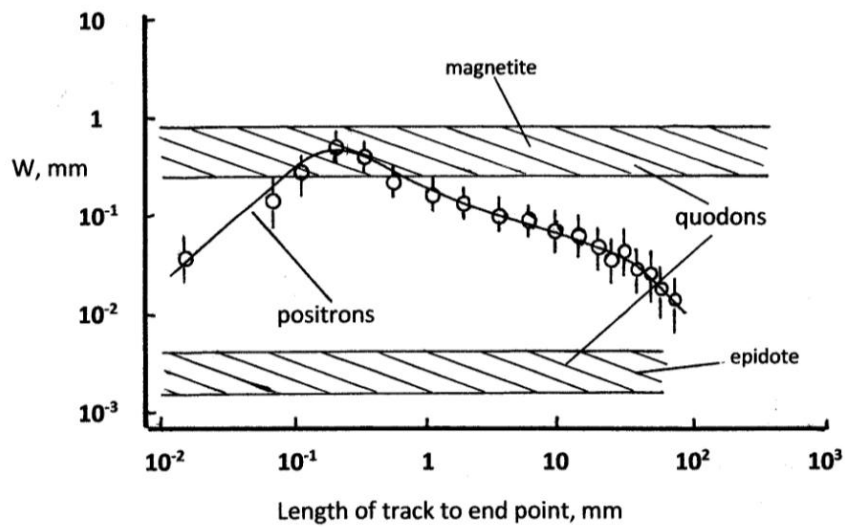

Figure 1. Plot showing the width of decorated positrons tracks as a function of distance from stopping point. The bands are the widths within which quodon tracks occur. The upper band is for decoration with magnetite and the lower with epidote.

The track of the positron is delineated with magnetite but the associated negative charge on the quodon inhibits the formation of magnetite but allows the formation of epidote. The width of tracks decorated with epidote is two orders of magnitude less than that for those with magnetite but is remarkably constant in width over the whole length of track and is the same on different tracks. This is consistent with an electron bound to the breather.

It has been shown that breathers can be created in nonlinear lattices by stochastic thermal processes[7,8,9]. Both stationary and mobile breathers are possible but the former are more selective on lattice parameters. Hence, it is reasonable to suppose that they can exist in layered high Tc materials as a sub-set of possible linear and nonlinear lattice excitations. Numerical studies of some high Tc materials have confirmed that the necessary conditions for breathers to exist and propagate do occur in one or more atomic planes in the layered crystals examined[10,11]. The present work suggests that at temperatures significantly above Tc nonlinear lattice excitations can bind securely to single charges. This raises the possibility that below Tc a second charge might bind to a quodon to form a Cooper pair. Such two-charge states have been proposed[12,13,14]. It has been shown that the Tc of all know HTSC materials can be related to just two lengths, one the length between holes in the conducting sheet and the other the lateral spacing between charges interacting in different sheets[1]. These lengths are comparable to the size of the 3D envelopes encompassing quodons. Finally, the stability of quodons at high temperature shows they are decoupled from phonons. This leads to the supposition that Cooper-pairs bound in a quodon will also be decoupled from phonons but can interact with other quodons with or without a bound charge. This suggests possible increase of Tc by reducing the energy of quodons propagating on intersecting chains.

Acknowledgements. I wish to thank J F R Archilla, J C Eilbeck and L Cruzeiro for helpful discussions.